\documentstyle[12pt]{article}
\begin{document}
\begin{center}
{\large\bf WORLDWIDE DEVELOPMENT OF ASTRONOMY}\\[0.2cm] 
{\bf THE STORY OF A DECADE OF UN/ESA WORKSHOPS ON BASIC SPACE
SCIENCE}\\[0.3cm]
{\large Hans J. Haubold}\\
UN Office for Outer Space Affairs\\
Vienna International Centre\\
Vienna, Austria\par
\smallskip
{\large Willem Wamsteker}\\
ESA IUE Observatory, VILSPA\\
Madrid, Spain\\
\end{center}
\bigskip
\noindent
{\bf Basic Space Science }\par
\smallskip
\noindent
\par
With the establishment of the United Nations Committee on the
Peaceful Uses of 
Outer Space$^1$ (COPUOS) in 1959, the Office for Outer Space
Affairs$^1$ of 
the United Nations Secretariat was assigned the responsibility
for implementing 
decisions of the Committee, its Legal Subcommittee and its
Scientific and 
Technical Subcommittee related to the promotion of
international cooperation in 
outer space matters. Among the primary tasks carried out by
the Committee is the 
development of international treaties, conventions or legal
principles governing 
the activities of Member States in the peaceful exploration
and use of outer space 
and the provision of technical assistance and information on
space science and technology to interested Member States.\par

Following the decision of COPUOS to take upon itself the
promotion of 
international cooperation in space science and technology, the
United Nations 
Programme on Space Applications$^1$ was established in 1969 with
the objective, 
inter alia, to provide scientists from developing countries
with educational 
programmes in remote sensing, satellite meteorology, satellite
communications, and 
basic space science. The activities of this programme are
subject to the annual review 
and approval of COPUOS. Because of the increasing importance
for developing 
countries to be actively engaged in space science research,
the United Nations and 
COPUOS have, from the beginning of the 1990s, begun to place an
increased emphasis on 
promoting education and research in space science and
technology, as well as 
planetary studies under the common denominator ``Basic Space
Science''.\par 

Of the 185  Member States of the United Nations, nearly 100
have professional or 
amateur astronomical organizations. Only about 60 of these
countries, however, 
support their astronomical community and scientific interests
through a membership 
in the International Astronomical Union (IAU) (Percy and
Batten$^2$). The 
distribution of first rank astronomical observatories has
shown strong change 
over the past 50 years, since many of the high quality
observing sites, required for 
the major facilities, are located in developing countries and
also the utilization of 
space observatories to overcome the limitations of the Earth
atmosphere, have 
introduced new parameters and supply previously unavailable
possibilities and 
mechanisms for the participation of developing countries in
Basic Space 
Science. \par
\medskip
\noindent
{\bf UN and ESA Initiated Workshops on Basic Space
Science}\par
\smallskip
In 1990, as part of the United Nations Programme on Space
Applications, the United 
Nations, in cooperation with the European Space Agency (ESA),
initiated the 
organization of annual Workshops on Basic Space Science\\
(Jasentuliyana$^3$) 
for developing countries as shown in Table 2. These Workshops 
were originally planned to be held as a 
unique series in each of the following five regions of the
world: Asia and the 
Pacific (ESCAP), Latin America and the Caribbean (ECLAC), 
Africa (ECA), Western Asia (ESCWA), and Europe (ECE). This 
subdivision by region follows the United Nations principles to
assess the 
relevance of space activities to worldwide economic and social
development.\par 
At the time of this writing, six UN/ESA Workshops on Basic
Space Science have been held, respectively in India
(1991), Costa Rica 
and Colombia (1992), Nigeria (1993), Egypt (1994), Sri
Lanka (1995), and Germany (1996) (cp. Table 2). 
An assessment of the 
achievements of this series of Workshops was made during
the sixth 
Workshop held in Bonn (Germany) in 1996; tthe region of Europe 
does not comprise developing countries but the Government of 
Germany made it possible to organize at least once such a 
Workshop in a highly industrialized country. The seventh Workshop 
will be held in Tegucigalpa, Honduras and there is an indication
of interest on part of the Government of Tunisia that the Workshop 
in 1998 will be organized in Tunisia.
Based on a decision of the United Nations General Assembly, the 
UNISPACE III Conference will be held at Vienna, Austria, in July 
1999 as a special session of COPUOS, open to all Member States 
of the United Nations. This conference will provide the 
opportunity to undertake a final assessment on what the UN/ESA 
Workshops on Basic Space Science have accomplished in making a 
contribution to the worldwide development of astronomy and 
space science.  In 1992, The
Planetary Society 
joined the organization of the UN/ESA Workshops in order to
establish a more 
equivalent emphasis on planetary science and exploration
in the programme of 
the Workshops. Other organizations joined in the support
on these UN/ESA 
Workshops, and taking into account the status in 1997,
the Workshops are co-organized by the German
Space 
Agency (DARA), the National Aeronautics and Space
Administration (NASA), 
the Institute for Space and Astronautical Sciences (ISAS), the
International Centre 
for Theoretical Physics (ICTP), the Austrian Space Agency
(ASA), and the French Space Agency (CNES) (Haubold 
et al.$^4$).\par Although the general content of the UN/ESA
Workshops was 
chosen to highlight the interest and importance of Basic
Space Science in a 
rather broad context, some special aspects were normally
addressing the interests 
of local organizers and the ongoing research activities in the
respective region.  
The topics included the following fields:
\begin{itemize}
\item International cooperation in basic space science;
\item Education for space science;
\item Solar terrestrial interaction;
\item Planetary science;
\item Space astronomy and astrophysics;
\item Cosmology; and
\item Data bases and on-line data in astronomy.
\end{itemize}

The Workshops were hosted by the Government of the respective
country in which 
they were held; Government representatives were actively
involved in the 
organizational and scientific preparations for the Workshops
which helped to 
enhance awareness and interaction between the Government and
the local 
scientific community.  This ultimately proved crucial to the
achievements of the 
Workshops' objectives.  A vital part of the programme of each
Workshop were the 
Working Group sessions which provide all participants with a
common
platform to 
make
critical observations and recommendations addressing the
development of Basic 
Space Science in their specific region. These observations
and recommendations 
are available in the published Workshop Proceedings$^{5-10}$
and are also 
contained in United Nations reports$^{11}$ on each
Workshop (carrying a UN documents number A/AC.105/...).  
These reports 
are available for distribution to governments (particularly
through COPUOS and 
its Scientific and Technical Subcommittee), national,
regional, and international 
organizations as well as concerned scientific institutions and
universities. This 
collection of observations and recommendations constitutes a
unique international 
framework for the development of astronomy covering five major
regions (and 
almost all developing countries) of the world, especially
highlighting the 
commonality of the problems and their possible solutions. To
prepare the 
Workshops on a regional basis, the United Nations invited
astronomers to submit 
studies on the current status of astronomy in the regions,
which will be published in a separate booklet titled 
``Developing Astronomy and Space Science Worldwide''.\par   
\medskip
\noindent
{\bf Follow-Up Projects Benefit Astronomy}\par
\smallskip
Although a number of large facilities for front-line research
are located in 
developing countries, these large, and often internationally
funded projects will 
not necessarily contribute to the overall scientific
development of the country 
where it is located. Unless a sufficiently qualified and
educated population exists 
in the host country, such institutes may not significantly
stimulate the overall 
participation in Basic Space Science. These facilities
take often mainly 
advantage of the climatological and geographic attributes of
any country in the 
world, particularly in developing countries, without
necessarily generating a 
culturally identifiable activity in the host country. On the
other hand it has been 
strongly emphasized that the establishment of networks of
existing small scientific 
facilities could be a much more efficient means to strengthen
international 
cooperation, particularly in geomagnetic studies, electrojet
current measurements, 
solar photometry, astrometry, galactic mapping, and optical
astronomy (the 1996 Workshop in Germany even addressed the 
question of the worldwide distribution of locations of 
facilities for neutrinos, gravitational-wave, and cosmic 
rays astronomy). Such 
Basic Space Science activities, which have their direct
interest associated with the 
climatological and socio-economic conditions of the developing
countries, can 
probably generate a considerably stronger support for the
Basic Space Science in 
the developing countries and may therefore be much more
efficient.\par
As an example the international, electronically-linked
observing program-\\
mes (similar to the `Whole Earth Telescope' project, the value of
which has  
become evident through successfully observing runs involving
tens of 
globally distributed telescopes in recent years), present
possibly very efficient 
means to stimulate the scientific development. Such programmes
could be expanded 
to include more active participation of developing countries
at relatively low cost 
to them.\par 
In addition to the common direct benefits of an international
scientific workshop, the 
UN/ESA Workshops have stimulated a number of follow-up
projects which have been implemented through the Workshops or will 
be implemented on a long-term basis.\par
\medskip
\noindent
{\it Astronomical Telescope Facility in Sri
Lanka$^{5,12,13}$}\par
\smallskip
The Government of Japan took the initiative to support the
establishment of 
national astronomical centers for astronomical education and
observational studies 
in the Asia and Pacific region (ESCAP) through the provision of
moderate-sized
astronomical research telescopes or planetaria. Over the past
several years, 
Singapore received a Mitaka-kokhi 40cm reflector for its
science centre of public 
education, and Malaysia started operating a Minolta
Planetarium at its space 
science education centre. Through the Japanese Cultural Aid
Grant Programme, 
Thailand was able to install a Goto 45cm reflector at the
Department of Physics of 
the Chulalongkorn University Bangkok, and at the Bosscha 
Observatory (Hidayat$^{14}$) in Lembang, Indonesia (the latter
 one is mainly used 
for astronomical research). As a result of the UN/ESA
Workshops, Sri 
Lanka has also received a 45cm astronomical telescope, which has
been  installed at the 
Arthur C. Clarke Centre for Modern Technologies at Katubedda
Moratuwa, Sri 
Lanka. Such moderate-sized optical telescopes, set up at
appropriate locations on 
Earth, can contribute in an important way to astronomical
research. The 45cm 
Goto telescope is equipped with a photometer, a spectrograph
and a photographic 
camera. Although the telescope was designed primarily for
photometric 
observational studies of variable stars, it also allows 
observations of comets and 
asteroids as well as studies of interstellar, interplanetary,
and atmospheric dust. A 
network of telescopes of this type, equipped with modern charge
coupled devices (CCD)
detectors and the 
appropriate small personal computer systems, throughout a
region, or even 
worldwide could form an very  powerful networked tool for many
types of 
astronomical research (Warner$^{10}$, Budding$^{10}$). Such a
 system
could foster 
regional and international cooperation in astronomical
research as in the case of 
the `spacewatch' programme.\par

\medskip
\noindent
{\it Colombia$^7$ and Honduras$^8$: Taking the Initiative in
Central
America}\par
\smallskip
An international cooperative programme for the Galactic Emission
 Mapping project (GEM) 
formed by Brazil, Colombia, Italy, and the United States of
America is
now operating a radio 
telescope in Colombia in order to survey the galactic emission
at long 
wavelengths (de Amici et al.$^7$). The GEM project is funded
by the United 
States National Science Foundation and by COLCIENCIAS in
Colombia.
Accurate and complete maps of the diffuse galactic emission in
the range of 0.5 - 10 GHz are required in order to study cosmic
ray
electrons in the galactic disk and the galactic magnetic field.
The galactic signal is also the most relevant foreground
contamination in cosmic microwave background radiation (CMBR)
experiments. Improving the understanding of galactic emission at
long wavelengths is therefore an essential task for extracting
cosmological information contained in present and future
generations of CMBR experiments. The Galactic Emission Mapping
project is carrying out observations from a number of sites
at different latitudes. The instrument consists of a 5.5-metre
parabolic reflector and receivers at 408, 2,300 and 5,000 MHz.
Taking data at several frequencies allows for the determination
of spectral indexes of the different emission processes.
Preliminary analysis and prototype tests indicated that a
substantial improvement over the existing maps will be achieved
within a few years from the beginning of the observing programme.
A first observing campaign from the White Mountain Research
Station in California, United States, has been completed in
November 1994. Observations from an equatorial site in Colombia
started in February 1995.
Because of its equatorial latitude and the presence of high
peaks in the Andean 
mountain range (higher than 4000 metres), Colombia is an ideal
location for astronomical
and radio 
astronomical research. Access to the northern and southern
celestial hemispheres 
with the same instrumentation opens a window of opportunity
for research in 
astronomy. Currently, scientific opportunities and the
feasibility of establishing an astronomical observatory in
Colombia are being pursued. 

The establishment of an Astronomical Observatory for Central 
America in  
Honduras, as well as the Central American Association of
Astronomers and Astrophysicists (Pineda de 
Carias$^8$), are very important for the
scientific development of 
the region as discussed extensively during the UN/ESA
Workshops on 
Basic Space Science. Building an astronomical
observatory for the six Central American countries has been
initiated in Honduras at the beginning of the 1990s. The 
 establishment of the observatory has already been
achived, following a strategy based on regional cooperation
involving national universities in Central America and at an
international level, by making contact with astronomers and
prestigious astronomical research centres. Since 1994, the
 astronomical observatory is operating at Tegucigalpa, at the
Universidad Nacional Autonoma of Honduras. This academic unit has
been equipped with a 42cm computerized telescope with a CCD and
other supporting facilities. In 1995, the observatory hosted the
first Central American course in astronomy and astrophysics and,
jointly with other European and Latin American universities, is
currently promoting a regional training programme for astronomers
of Central America. Several important cooperation agreements are
in the process of being signed in order to contribute to the
development of Basic Space Science in the region. The observatory 
will host the seventh UN/ESA Workshop on Basic Space Science in 
June 1997 (cp. Table 2).

\medskip
\noindent
{\it A Large Astronomical Facility for Africa?$^8$}\par
\smallskip
A proposal for an Inter-African Astronomical Observatory and
Science Park on 
the Gamsberg in Namibia was endorsed by the
United Nations as a 
result of the UN/ESA Workshops. Because of its unique
geographic location, 
southern Africa can make an immense contribution to Basic
Space Science. 
For time-critical phenomena in astronomy, 24-hour coverage can
only be obtained
through astronomical observatories in continents (excluding
Antarctica) south of 
the equator. The Gamsberg has been identified as one of the
most suitable sites for 
an observatory in southern Africa. It is a table mountain
120km south-west of 
Windhoek above the Namib desert at an altitude of 2350m above
sea level. It 
experiences a large number of cloudless nights, a dark sky,
excellent atmospheric
transparency and low humidity, equal to the well-known
astronomical sites in 
Chile. The mountain top is owned by the Max-Planck-Society of
Germany and a 
small astronomical station was established there in the 1970s
(Elsaesser$^8$). 
Besides astronomy, the mountain is of considerable interest to
other scientific 
disciplines such as cosmic ray physics, atmospheric research
and meteorology. 
The huge plateau of about 250 hectares offers enough space for
various  
independent installations. The Max-Planck-Institute for
Astronomy at Heidelberg, 
Germany, made efforts to initiate the development of a new
scientific centre on the 
Gamsberg. This can only be achieved, however, with strong 
international collaboration 
and support in kind and cash. South Africa had expressed an
interest in operating 
the astronomical observatory on behalf of the international
community. The ideal 
solution would be an Inter-African Science Park. The
Government of Namibia, as 
well as the Windhoek University, had also 
expressed support for 
this project. If established in the future, this facility 
could become an important focus in the 
development of Basic Space Science in African countries (Okeke
and 
Onuora$^8$). If the facility is established with a viable
infrastructure, it is 
possible that it would be attractive to northern hemisphere
countries as well, 
especially those wishing to establish facilities in the
southern hemisphere. The United Nations has been informed in 
1996 that current circumstances do not allow to further pursue 
the project. \par 
\medskip
\noindent
{\it The Kottamia Observatory: The Largest in Western
Asia$^{9,10}$}
\par
\smallskip
The Kottamia Observatory, housing a 74-inch reflecting
telescope, is located in 
the north-eastern Egyptian desert, 80km from Helwan, Egypt, on
the Cairo-Suez 
road. Because the Kottamia Observatory telescope facility
can supply major 
observational capabilities for Basic Space Science in the
region, its importance 
cannot be underestimated. Hence, to safeguard the importance
of this telescope 
facility, further
development was needed 
(Mikhail$^{9,10}$). Mindful of  this, deliberations of the
UN/ESA Workshops 
noted with great satisfaction that there is a willingness to
cooperate on a regional basis in this effort.
This has in the meantime resulted in a more formal
collaboration between 
scientists in the Western Asia region to enhance  the Kottamia
Observatory and at 
the same time share the other existing astronomical facilities
in the region. This 
represents a very important new scientific collaboration
initiative
in Western Asia. It is the 
intention of the Government of Egypt to open the Kottamia
Observatory to 
scientists from the entire region in order to allow access to
the observational 
opportunities for Basic Space Science. The plans of
astronomers from other 
countries, both in Western Asia and Africa, to share in the
future
development of the Kottamia Observatory, will 
contribute to the future 
development of Basic Space Science in the region and will
enhance significantly 
the opportunities for cooperation among astronomers in the
region. It was finally decided in 1994 to contract the
refurbishment of
the Kottamia telescope after the evaluation of several submitted
tenders. The National Research Institute of Astronomy and
Geophysics (NRIAG) at Helwan and the Ministry for Science and
Education of Egypt entered into a contract financed completely by
the Egyptian Government. The task includes the design and
manufacture of a new optical system for the 74-inch telescope
tube. The mirror materials are made from Schott Zerodur to ensure
superb optical quality in the respective temperature range for
observations. In order to achieve a high-quality optical surface
in working conditions, i.e. in all applicable positions of the
telescope, a new support or mirror cell for the primary mirror
was necessary. Therefore a new 18-point support instead of
the old 9-point support was proposed and was part of the
project. The new optics is integrated in the nearly
30-year-old Kottamia telescope and first light is expected in June 
1997. In July 1995 the representatives of NRIAG accepted the test
results of the blanc for the primary mirror at Carl Zeiss in
Germany. The mirror is still being ground and polished, resting
on an 18-point support just as in the future telescope cell. The
procedure has taken several months, first creating a spherical
surface of already high surface quality and then gradually
approximating the required spherical shape. Preliminary tests of
the mirror shape showed excellent results, and the preliminary
acceptance tests were accomplished according to schedule in
1996. As part of this 
collaboration also plans are under consideration for the
identification of a high 
quality astronomical site in the region for completely new
state-of-the-art 
modern large astronomical telescopes.\par 
\medskip
\noindent
{\bf Participation of Egypt in Future Mars Rover Mission}\par
\smallskip
The Planetary Society, a major cosponsor of the UN/ESA Basic Space
Science Workshops, is following up the suggestion that Egypt
participate in the US-Russia Mars Rover mission in 2001 through
involvement in the design, building and testing of a drill for
obtaining subsurface samples. The Planetary Society informed the
Space Research Institute (IKI) of the Russian Academy of Sciences
about the idea, and they, in turn, formally invited the Egyptian
Ministry of Scientific Research to study the concept for
potential use on the US-Russia Mars 2001 mission. Of the many
important scientific objectives of the Marsokhod mission, among
the most interesting, is the analysis of sub-surface samples.
Inclusion of some sort of drilling mechanism in the payload of
such a mission would assist scientists in the investigation of
volatiles, organic materials and mineralogy. Twenty years ago,
the arm on the Viking Mars lander was able to obtain samples from
depths up to 10 cm. Today, a drill with the capability of boring
at least an order of magnitude deeper (more than one metre) would
be essential to further research and investigation. Egypt has
expertise in drill development. Many years ago, as part of the
archaeological exploration of the pyramids, a sophisticated
drilling system was developed to drill into and deploy a camera
into a sub-surface chamber without allowing air into the chamber.
The drill perforated the limestone to a depth of 2m without the
use of lubricants or cooling fluids that might have contaminated
the pit's environment, and successfully collected (six) samples.
This experience as well as more common terrestrial applications
suggest that the necessary technology base for a drill
development can be brought together. In the proposed application
for the Mars 2001 mission, Egypt would assume the financial
responsibility for the drill as part of their Marsokhod
participation. A study team of Egyptian scientists, collaborating
with American, Russian, and European scientists, is now pursuing
the project. \par
\medskip
\noindent
{\bf World Space Observatory}\par
\smallskip
In all UN/ESA Workshops on Basic Space Science, it was stated
that, considering the increase in the participation
of the developing countries in astronomy and space science and
taking into account the foreseeable rapid increase of
participating professionals in the developing countries, it is
important to establish the tools for their participation at the
most advanced level. Since access to smaller telescopes and the
use of archival data in astronomy would result in an expanding
and professionally competent astronomical community in the
developing countries, it should be recognized that access to
front-line facilities are required for many scientists. As the
costs associated with major ground-based facilities often pose
excessive economic burdens for the developing economies, such
conditions give rise to an unproductive conflict cycle in which
many of the best trained scientists tend to travel elsewhere for
their professional lives, which would remove an important asset
for their countries: highly trained people. In a world where
concentration of first-scale astronomical facilities is an
unstoppable trend, a technologically attractive solution could be
supplied by a World Space Observatory. That would also stimulate
industrial development, enhance and improve the communications
infrastructure and allow independent local access to a prime
astronomical facility.\par
\medskip
\noindent
{\bf A Worldwide Network of Telescopes for Observing
Near-Earth Objects? 
$^{15,16}$}\par
\smallskip
The recent impact of the comet Shoemaker-Levy 9 on Jupiter has
renewed the 
interest in the small bodies in the inner solar system ``Near
Earth Objects'' 
(NEOs). Pursuing the understanding of the Near-Earth Objects
has become an issue 
of global interest. The Explorers Club and the United Nations
Office for Outer 
Space Affairs organized an International Conference
on Near-Earth 
Objects (Remo$^{15}$).
Researchers in the fields of astronomy, planetary science,
astrophysics, 
paleontology and astronautics joined in a multi-disciplinary
forum to discuss 
related topics. Among the topics addressed by the conference
was the 
establishment of observational facilities dedicated to NEO
studies. As a first step, 
these facilities could be associated with existing small
astronomical observatories, 
also in developing countries. Observational programmes could be
coordinated with 
activities of amateur astronomy groups and organized on an
international scale  
which may lead to the establishment of a network of
moderate-sized astronomical 
telescopes as discussed in the UN/ESA Workshops on Basic Space
Science.\par
\medskip
\noindent
{\bf Usage of Archival Data}\par
\smallskip
As a consequence of the UN/ESA Workshops, scientists have also
become much 
more aware of the importance and availability of archival
data, especially from 
space observatories, and their importance for educational
purposes as well as the 
capabilities to do front-line science even with modest means.
One of the most 
widely spread archives of space data is the archive of the
International Ultraviolet 
Explorer satellite (IUE). The concept and design of its
distribution system ULDA 
(Uniform Low Dispersion Archive) has made the distribution of
these data to 
many sites where no previous experience with space data could
be foreseen a 
practical reality (Wamsteker et al.$^{17}$). Figure 1 
illustrates the worldwide 
distribution of the IUE/ULDA. To show how the usage of the
data in the 
developing countries compares with that in the ESA Member
States, Table 1 show the usage statistics over the past 8 years. 
It becomes evident from these data that these facilities 
are an important research tool also for scientists in developing 
countries.

\medskip
\noindent
{\bf Conclusions}\par
\smallskip
The observations and recommendations made during the past six
 UN/ESA 
Workshops on Basic Space Science can be summarized in the
following topics 
which need to be addressed urgently on a regional and
international level: (i) 
promotion of the advancement and dissemination of the
knowledge of Basic Space
Science and its application to human welfare, (ii) provision of
on-line databases
and electronic mail services, 
(iii) provision of abstracting and indexing services in Basic
Space Science, (iv) 
dissemination of reliable information on Basic Space Science
to the public, (v)
collection and analysis of the statistics on the profession
and on education in Basic 
Space Science, (vi) encouragement of the documentation and
study of the history 
and philosophy of Basic Space Science, and (vii) cooperation
with national, regional, and international organizations on
educational projects at all levels.\par
Among the above topics, the electronic networking of
scientific institutions may 
have the most immediate impact on the situation in developing
countries. There 
exist large data archives in space science readily
available at virtually no cost 
to any astronomer which has established access to the
INTERNET and the World Wide Web. Space 
astronomy missions such as IUE, HST, COBE, ROSAT, IRAS, etc.,
have
made their 
data archives publicly accessible through electronic 
networks.$^{10,17}$
These archives are available to astronomers in any country on
Earth as long as 
they have access to the INTERNET and the World Wide Web. These 
electronic networks
also allow 
immediate access to electronic mail channels and electronic
publications 
(e.g. the Astrophysical Data System), solving the traditional problem
of isolation and 
obsolete libraries in many developing countries. The combined
efforts of 
individual astronomers and the support of governments and
international 
organizations could realize the
concept of a ``Global Village'' in terms of astronomy and space science 
education and
research worldwide.\par 
\clearpage
\noindent
{\bf References}\par
\smallskip
\noindent
1.  1982. The World in Space. R.Chipman, Ed. \par
Prentice-Hall, Inc. Englewood  Cliffs, New Jersey.\\
2.  Percy,J.R. and Batten,A.H. 1995. Mercury 24:2, 15-18.\\
3.  Jasentuliyana,N. 1995. Space Policy 11:89-94.\\
4.  Haubold,H.J., Ocampo,A.C., Torres,S., and Wamsteker,W.
1995. \par
ESA Bulletin, February, No.81,p.18-21.\\
5.  1992. Basic Space Science. H.J.Haubold and R.K.Khanna,
Eds.\par 
American Institute of Physics Conference Proceedings, Vol.
245,\par
AIP, New York.\\
6.  1993. Basic Space Science. W.Fernandez and H.J.Haubold,   
Eds.\par 
Earth, Moon, and Planets 63:93-170.\\
7.  1994. Basic Space Science. H.J.Haubold and S.Torres,
Eds.\par  
Astrophysics and Space Science 214:1-260.\\
8.  1995. Basic Space Science. H.J.Haubold and L.I.Onuora,
Eds.\par 
American Institute of Physics Conference Proceedings,
Vol. 320,\par 
AIP, New York.\\
9.  1995. Basic Space Science. H.J.Haubold and J.S.Mikhail,
Eds.\par 
Earth, Moon, and Planets 70:1-229.\\
10.  1996. Basic Space Science. H.J.Haubold and J.S.Mikhail,
Eds.\par 
Astrophysics and Space Science 228:1-405.\\
11. Brief reports on the UN/ESA Workshops and the status of the\par 
follow-up projects are available on the World Wide Web at\par 
the electronic address\par                          
http://ecf.hq.eso.org/$\sim$ralbrech/un/un-homepage.html.\\
12. Gehrels,T. 1988. On the Glassy Sea: An Astronomer's Journey.\par 
American Institute of Physics, New York.\\
13. Wickramasinghe,C. 1994. Ed. Fundamental Studie\par 
and Future of Science. University College Cardiff\par 
Press, Cardiff, pp. 377-385.\\
14. Hidayat,B. et al. 1992. In Evolution of Stars and\par 
Galactic Structure. K.Isida and B.Hidayat, Eds.\par 
National Astronomical Observatory. Tokyo.\\
15. Remo,J.L. 1996. Space Policy 12:13-17.\\
16. 1992. The Spaceguard Survey: Report of the NASA
International\par 
Near-Earth Object Detection Workshop. D.Morrison, Ed.\par 
National Aeronautics and Space Administration, Washington.\\
17. Wamsteker,W., et al., 1989, IUE-ULDA/USSP:\par 
The on-line data archive (low resolution) of the \par
International Ultraviolet Explorer,.\par 
Astron. Astrophys. Suppl. Ser., 79, 1.\\
\begin{tabular}{|l|l|l|l|}
\multicolumn{4}{c}{\bf Table 2. Overview on the series of Basic Space Science Workshops}\\[0.5cm] \hline
{\bf Year} & {\bf City} & {\bf Target} & {\bf Host}\\
& {\bf Country} & {\bf Region} & {\bf Institution} \\ \hline
1991 & Bangalore, India & Asia and the Pacific & Indian Space Research \\
& & ESCAP & Organization (ISRO)\\ \hline
1992 & San Jos\'{e}, Costa Rica & Latin America and & University of Costa Rica\\
& Bogota, Colombia & the Caribbean & University of the Andes\\
& & ECLAC & \\ \hline
1193 & Lagos, Nigeria & Africa & University of Nigeria \\
& & ECA & and Obafemi Awolowo \\ 
& & & University \\ \hline
1994 & Cairo, Egypt & Western Asia & National Research \\
& & ESCWA & Institute of Astronomy \\
& & & and Geophysics (NRIAG)\\ \hline
1995 & Colombo, Sri Lanka & Asia and the Pacific & Arthur C. Clarke\\
& & & Center for Modern\\
& & & Technologies (ACCMT)\\\hline
1996 & Bonn, Germany & Estern and Western & Max-Planck-Institute\\
& & Europe  & for Radioastronomy\\
& & ECE & (MPIfR)\\ \hline
1997 & Tegucigalpa, Honduras & Central America & Astronomical Observatory \\
& & & of the Autonomous National\\
& & & University of Honduras (OA/UNAH)\\ \hline
1998 & Tunis, Tunisia & Africa &\\ \hline
1999 & Vienna, Austria & UNISPACE III Conference & UN COPUOS \\ \hline
\end{tabular} 
\clearpage
\begin{tabular}{|l|c|l|l|}
\multicolumn {4}{c} {\bf Table 2. Continued} \\[0.5cm] \hline 
{\bf Year} & {\bf Number of} & {\bf Topic/sub-topic of} & {\bf Follow-up project/UN Report}\\
& {\bf Participants/} & {\bf the Workshop} & \\ 
& {\bf countries} & &\\ \hline
1991 & 87 & Basic Space Science & Establishment of Astronomical Facility at\\
&19 & & Sri Lanka Recommended A/AC.105/489\\ \hline
1992 & 122 & Basic Space Science & Establishment of Astonomical Observatory \\
& 19 & & for Central America Recommended \\ 
& & & Establishment of Radiotelescope in Colombia\\
& & & Recommended A/AC.105/530\\ \hline
1993 & 54 & Basic Space Science & Establishment of Inter-African Astronomical\\
& 15 & & Observatory and Science Park at Namibia \\
& & & Recommended A/AC.105/560/Add.I\\ \hline
1994 & 95 & Basic Space Science & Refurbishment of Kottamia Telescope\\
& 22 & & Recommended\\
& & & Participation of Egypt at US/Russia Mars\\
& & & Mission 2001 Recommended A/AC.105/580\\ \hline
1995 & 74 & From Small & Inauguration of Astronomical Facility at\\
& 25 & Telescpes to Space & Sri Lanka \\
& & Missions & A/AC.105/640\\ \hline
1996 & 120 & Ground-based and & Assessment of the Achievements of the Whole \\
& 34 & Space-borne & Series of UN/ESA Workshops \\
& & Astronomy & Foundation of Working Group on Basic Space\\
& & & Science in Africa  A/AC.105/657\\ \hline
1997 & 75 & Small Astronomical & Inauguration of Astronomical Observatory for\\
& 25 & Telescopes and & Central America at Honduras\\
& & Satellites in & \\
& & Education and & \\
& & Research & \\ \hline
1998 & & & \\ \hline
1999 & open to all & &\\
& Member States & \\ \hline
\end{tabular}
\end{document}